\documentclass[sn-mathphys,Numbered]{sn-jnl}

\usepackage{graphicx}
\usepackage{multirow}
\usepackage{amsmath,amssymb,amsfonts}
\usepackage{amsthm}
\usepackage{mathrsfs}
\usepackage[title]{appendix}
\usepackage{xcolor}
\usepackage{textcomp}
\usepackage{manyfoot}
\usepackage{booktabs}
\usepackage{algorithm}
\usepackage{algorithmicx}
\usepackage{algpseudocode}
\usepackage{listings}
\usepackage{comment}

\theoremstyle{thmstyleone}

\theoremstyle{thmstyletwo}

\theoremstyle{thmstylethree}

\begin{document}

\title[Scaling law of Sim2Real transfer learning in materials science]{Scaling Law of Sim2Real Transfer Learning in Expanding Computational Materials Databases for Real-World Predictions}

\author*[1]{\fnm{Shunya} \sur{Minami}}\equalcont{These authors contributed equally to this work.}
\author[1,2]{\fnm{Yoshihiro} \sur{Hayashi}}
\equalcont{These authors contributed equally to this work.}
\author[1,2]{\fnm{Stephen} \sur{Wu}}
\author[1,2]{\fnm{Kenji} \sur{Fukumizu}}
\author[3]{\fnm{Hiroki} \sur{Sugisawa}}
\author[4]{\fnm{Masashi} \sur{Ishii}}
\author[4]{\fnm{Isao} \sur{Kuwajima}}
\author[3]{\fnm{Kazuya} \sur{Shiratori}}
\author*[1,2,4]{\fnm{Ryo} \sur{Yoshida}}\email{yoshidar@ism.ac.jp}
\affil[1]{\orgdiv{The Institute of Statistical Mathematics}, \orgname{Research Organization of Information and Systems}, \orgaddress{\city{Tachikawa}, \postcode{190-8562}, \country{Japan}}}
\affil[2]{\orgdiv{The Graduate Institute for Advanced Studies}, \orgname{SOKENDAI}, \orgaddress{\city{Tachikawa}, \postcode{190-8562}, \country{Japan}}}
\affil[3]{\orgdiv{Science \& Innovation Center}, \orgname{Mitsubishi Chemical Corporation}, \orgaddress{\city{Yokohama}, \postcode{227-8502}, \country{Japan}}}
\affil[4]{\orgdiv{Research and Service Division of Materials Data and Integrated System}, \orgname{National Institute for Materials Science}, \orgaddress{\city{Tsukuba}, \postcode{305-0047}, \country{Japan}}}

\abstract{
To address the challenge of limited experimental materials data, extensive physical property databases are being developed based on high-throughput computational experiments, such as molecular dynamics simulations.
Previous studies have shown that fine-tuning a predictor pretrained on a computational database to a real system can result in models with outstanding generalization capabilities compared to learning from scratch. 
This study demonstrates the scaling law of simulation-to-real (Sim2Real) transfer learning for several machine learning tasks in materials science. 
Case studies of three prediction tasks for polymers and inorganic materials reveal that the prediction error on real systems decreases according to a power-law as the size of the computational data increases.
Observing the scaling behavior offers various insights for database development, such as determining the sample size necessary to achieve a desired performance, identifying equivalent sample sizes for physical and computational experiments, and guiding the design of data production protocols for downstream real-world tasks.
}

\maketitle

\section*{Introduction}\label{sec1}

Machine learning holds great potential for revolutionizing the methodology of materials science. Recent studies have demonstrated that models trained using materials data can accurately predict various physicochemical properties for diverse material systems \citep{yamada2019predicting, aoki2023multitask}. Conventionally, a model defines the mapping from compositional or structural features of a given material to its thermal, electrical, mechanical, and energetic properties, as well as higher-order structural features. Assessing a large library of candidate materials using such models has led to the discovery of various materials, such as
polymers \citep{wu2019machine}, inorganic crystalline compounds \citep{merchant2023scaling, szymanski2023autonomous}, high-entropy alloys \citep{rao2022machine}, catalysts \citep{zhong2020accelerated, kim2019artificial}, and quasiperiodic materials
\citep{liu2021machine, liu2023quasicrystals, uryu2024deep}. The success of such data-driven research depends on the quantity and quality of the data, and researchers often face the critical issue of data scarcity. Generating experimental data requires time-consuming, multi-stage workflows involving synthesis, sample preparation, property measurements, phase identification, and other laborious trial-and-error processes. More critically, researchers lack the incentive to disclose their laboratory data to open communities due to concerns regarding information confidentiality \citep{Martin2023-xd}, which hampers the co-creation of an open data foundation.

Large-scale databases based on computer experiments such as first-principles calculations and molecular dynamics (MD) simulations are being developed to overcome the barriers posed by limited experimental data. For inorganic compounds, extensive first-principles property databases, including tens of thousands or more of crystal structures, have been developed, such as Materials Project \citep{Jain2013}, AFLOWLIB \citep{curtarolo2012aflow}, NOMAD \citep{draxl2019nomad}, OQMD \citep{saal2013materials, kirklin2015open}, and GNoME \citep{merchant2023scaling}. The QM9 database \citep{ramakrishnan2014quantum} comprises over 130,000 small organic molecules, providing molecular structures and their properties obtained from quantum mechanical calculations, which serves as a dataset for machine-learning-based property prediction tasks \citep{wu2019machine, yamada2019predicting}. 
Although there is currently no comprehensive computational database for polymeric materials, RadonPy \citep{hayashi2022radonpy} is being developed as a Python library for fully automated all-atom classical MD simulations to generate data resources for machine learning.

The methodology of transfer learning, particularly simulation-to-real (Sim2Real) transfer, enables the integration of extensive simulation data with limited quantitative experimental data \citep{su2015render, movshovitz2016useful, georgakis2017synthesizing, tremblay2018training}. Transfer learning is beneficial when training a model from scratch on the target task is impractical due to data scarcity; it leverages data or pretrained models from a source domain to enhance machine learning tasks in a target domain. This becomes increasingly advantageous as the relevance between the source and target domains increases. For example, in computer vision, Sim2Real transfer is crucial for adapting vision models trained in simulation environments to real-world applications, such as autonomous vehicles, by leveraging insights gained from simulated environments. Sim2Real transfer is also widely used in materials research. For instance, \citet{wu2019machine} developed a predictive model for the thermal conductivity of polymeric materials using experimentally observed data for 28 amorphous polymers. Leveraging a large dataset of specific heat capacity generated through quantum chemistry calculations as the source task, they successfully derived the Sim2Real-transferred model in the target domain. Similarly, \citet{aoki2023multitask} employed a machine learning framework called multitask learning to integrate a quantum chemistry dataset with biased and quantitatively limited experimental data, successfully building a predictive model for polymer--solvent miscibility for a wide range of chemical spaces. \citet{ju2021exploring} employed transfer learning to build a model predicting lattice thermal conductivity of inorganic crystalline materials. With only 45 samples for the target property, ordinary supervised learning failed to meet accuracy requirements. To address this, they utilized the first-principles calculations of scattering phase space as the source task and applied transfer learning to achieve sufficient accuracy.

Given the inherent domain gap between computer experiments and real-world systems, it is uncertain whether increasing the volume of simulation data enhances the generalization performance of Sim2Real-transferred models. To clarify this, \citet{Mikami2021ASL} provided theoretical and experimental evidence showing that the generalization of Sim2Real transfer learning improves according to a power-law relationship with the expansion of simulation data. Specifically, experimental validation of the scaling law was demonstrated for the Sim2Real scenario in computer vision tasks. Observing the scaling behavior of Sim2Real transfer, and estimating its convergence rate and asymptotic behavior offer valuable insights for advancing database development. 

Here, we present a statistical measure for quantitatively evaluating the transferability and scalability of a growing computational materials database. Our work reveals the existence of a scaling law in transfer learning across diverse prediction tasks in materials research involving polymers and inorganic material systems. Specifically, it encompasses three scenarios: (1) Sim2Real prediction of polymer properties by re-purposing neural networks pretrained with all-atom classical MD simulations; (2) multitask machine learning integrating expansive data from quantum chemistry calculations and a limited experimental dataset to predict the miscibility of polymer--solvent binary mixtures; and (3) valification of the Wiedemann--Franz (WF) law between thermal and electrical conductivities of inorganic material systems through transfer learning. Notably, since both the source and target datasets in the third case were obtained from real experiments, the concept shown here can extend beyond Sim2Real scenarios. By experimentally observing the scaling behavior of transferred predictors, we can estimate their expected generalization performance upon further increasing the volume of simulation data, serving as an indicator of the database’s potential value. Moreover, multidimensional scaling, considering both physical and computer experiments, provides a statistical estimate for the equivalent sample size of experimental and computational data. This aids in decision-making for the design of data production protocols. Additionally, by observing the scaling behavior of individual materials, we can individualize database design guidelines and gain insights into the existence of material groups that share physical mechanisms across different material systems. 

\section*{Results}\label{sec2}

\subsection*{Outline}

Sim2Real transfer learning involves adapting a predictive model that is pretrained in a virtual environment to real-world scenarios. In materials research, the predictor defines a mathematical mapping from a descriptor representing the composition or structural features of a given material to its physicochemical properties. In the source task, the model is trained using a dataset of size $n$ generated from computer experiments, such as first-principles ab initio calculations or MD simulations. In the target task, this pretrained model is repurposed and transferred to predict experimentally observed properties, utilizing an experimental dataset of size $m$, where $m$ is typically much smaller than $n$. \citet{Mikami2021ASL} presented a general theory, under certain assumptions, stating that in the fine-tuning of neural networks, the generalization error $\mathbb{E}[L(f_{n,m})]$ with the squared loss $L(f)$ of a transferred model $f_{n,m}$ for the real-world system is bounded from above by a function 
$R(n, m)$:
\begin{eqnarray}
R(n,m) := (A n^{-\alpha} + B )m^{-\beta}  + \epsilon, \label{eq:scale_nm}
\end{eqnarray}
where $A, B, \alpha, \beta, \epsilon \ge 0$ are constants independent of $n, m$.
In particular, considering the case of a fixed number of experimental samples at $m$, the upper bound for the generalization error is expressed as follows:
\begin{eqnarray}
\mathbb{E}[L(f_{n,m})] \le R(n) := D n^{-\alpha} + C, \label{eq:scale_n}
\end{eqnarray}
where $D:=A m^{-\beta}$ and $C:=B m^{-\beta} + \epsilon$.
According to this law, as $n$ increases, the generalization error of predicting experimentally observed properties for the transferred network converges to a reachable limit $C \ge 0$, called the transfer gap, with a decay rate $\alpha \ge 0$.
\citet{Mikami2021ASL} demonstrated that these power-law relations hold empirically in Sim2Real transfer in computer vision tasks.

While increasing the data size for pretraining, the reduction in the generalization error of the transferred model is measured experimentally, which can be used to evaluate whether the transferred model can attain the desired prediction performance in the target task or to estimate the required sample size based on the estimated $(D, \alpha, C)$. Additionally, the observed scaling behaviors provide guidelines for designing the source database. For a predefined set of downstream tasks leveraging the database, the simulation environment can be tailored to accelerate scaling to real systems, such as selecting empirical interatomic potentials or polymerization degrees in MD simulations. For instance, \citet{Mikami2021ASL} applied Sim2Real transfer for computer vision tasks and showed that intentionally increasing the diversity of the appearance, luminosity, and background in a synthetic image set leads to an increase in the scaling factor $\alpha$ and a partial improvement in $C$. Creating a data-generation scheme that results in a negligible $C$ is the ultimate objective in developing foundational source data. Intuitively, the consistency of simulations to real-world scenarios and the methodology employed in transfer learning mainly affect the magnitude of $C$. 

In the following sections, we describe the benefits and utility of analyzing scaling laws in transfer learning, based on three case studies of different material systems and their databases.

\subsection*{Polymer property predictions with Sim2Real transfer learning using MD simulations}
\label{sec:MD}

We demonstrate that the scaling law of Sim2Real transfer learning holds in polymer property prediction using all-atom classical MD simulations. The target properties to be predicted are the refractive index, density, specific heat capacity at constant pressure ($C_{\mathrm P}$), and thermal conductivity. Using RadonPy \citep{hayashi2022radonpy}, an open-source Python library developed to fully automate MD simulations of polymeric materials using LAMMPS (large-scale atomistic/molecular massively parallel simulator) \citep{thompson2022lammps}, we generated a source dataset comprising the four physical properties of approximately $7 \times 10^4$ amorphous polymers (see Table S1 for the number of samples). Details of the MD calculations are provided in the Methods section.
We randomly selected $n$ samples from this dataset for the pretraining of neural networks, where $n$ was varied across 10 equally spaced points on a logarithmic scale between $100$ and the maximum number of samples.

\begin{figure}[ht]
    \centering
    \includegraphics[clip, width=\columnwidth]{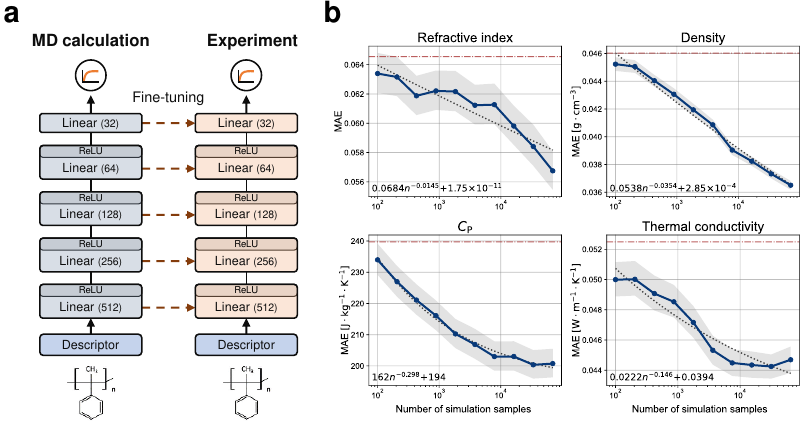}
    \caption{Transfer learning of polymer property predictions using all-atom classical MD simulations. (a) Neural network architecture. (b) Scaling behavior of Sim2Real transfer for four different properties, namely refractive index, density, specific heat capacity ($C_{\mathrm P}$), and thermal conductivity. The horizontal axis represents the simulation data size, and the vertical axis shows the MAE averaged over 100 independent trials with 90\% confidence interval calculated by performing bootstrapping sampling. The dashed line is the estimated power-law with the estimated equation given at the bottom left, and the horizontal red line indicates the mean MAE for direct learning with no pretraining.}
    \label{fig:MD}
\end{figure}

The property predictor used a 190-dimensional descriptor vector that represents the compositional and structural features of the chemical structure of a polymer repeating unit. This vectorized polymer was mapped to each property using a conventional fully connected multi-layer neural network (see Fig.\ref{fig:MD}a and the Method section). With experimental data, we fine-tuned each pretrained neural network to a predictor of the experimental properties. The experimental datasets were extracted from the PoLyInfo database \citep{polyinfo_paper, polyinfo_paper2, polyinfo_web}. The number of polymers in each property dataset was 234 for refractive index, 607 for density, 104 for $C_{\mathrm P}$, and 39 for thermal conductivity. To transfer a pretrained model to each target domain, we randomly selected 80\% of the experimental datasets and evaluated the model’s predictive performance on the remaining samples. This process was repeated 500 times independently for each $n$, observing scaling behaviors with the average of the mean absolute errors (MAEs) with their 90\% confidence interval calculated by performing bootstrapping sampling.

As shown in Fig. \ref{fig:MD}b, the empirical generalization error for the experimental refractive index decays almost linearly on a logarithmic scale across the observed range of $n$. The parameters for power-law scaling were estimated as $D=0.0684$, $\alpha=0.0145$, and $C=1.75 \times 10^{-11}$.  
For the density, the prediction error also linearly decreases, and as $n$ grows infinitely large, the MAE is expected to approach zero. The prediction error for $C_{\mathrm P}$ linearly decreases until around $n=10^4$, after which the decay begins to slow. Regarding the thermal conductivity, the generalization performance rapidly improves until $n < 10^4$, followed by a plateau as $n$ further increases. In summary, all tasks are notably scaled as the volume of MD-calculated data increases. Moreover, as shown in Fig. \ref{fig:MD}b, the generalization performance of transfer learning notably surpasses that of direct learning without transfer. The potential cross-domain transferability becomes more evident when contrasting direct and transfer learning based on scaling behavior rather than at a fixed $n$.

For the refractive index and density, the MD-calculated values exhibit remarkably high consistency with the experimental observations from our previous study \citep{hayashi2022radonpy}. Therefore, the observed strong scaling is verified because increasing the amount of simulation data directly improves the generalization performance for real-world scenarios. The $C_\mathrm{P}$ calculations with the classical MD simulations (neglecting quantum effects) introduced systematic biases compared to the corresponding experimental values, resulting in significant overestimation of the $C_\mathrm{P}$ \citep{hayashi2022radonpy}. 
Furthermore, the effect of random sampling of the initial structures in the MD simulations is more pronounced for the MD-calculated $C_{\mathrm{P}}$ than for the refractive index and density. Similarly, there are slight systematic biases and inherently large fluctuations in the MD-calculated thermal conductivity values. Hence, these findings suggest that simulation uncertainty due to the randomness of initial structures is one of the critical factors influencing the scaling strength. This aspect will be discussed in more detail later.

\begin{figure}[ht]
    \centering
    \includegraphics[clip, width=\columnwidth]{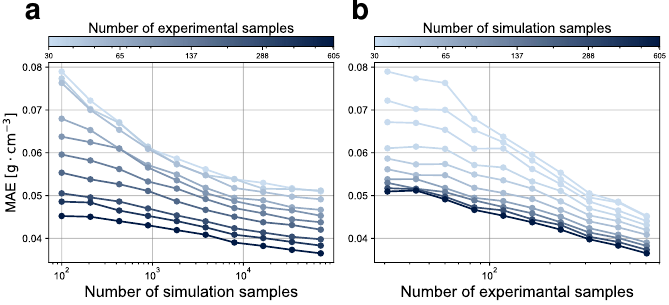}
    \caption{
        Multidimensional scaling of Sim2Real transfer learning, illustrated by the density prediction of amorphous polymers. (a) Scaling to increase the amount of simulation data across various experimental dataset sizes, and (b) scaling to increase the amount of experimental data for different sizes of simulation datasets. Each line represents the MAE averaged over 500 independent trials.
    }
    \label{fig:multiscale_radonpy}
\end{figure}

Here, we discuss the multidimensional scaling of simulated and experimental data. We examined the scaling behavior of Sim2Real predictions for the density while simultaneously varying the sizes of simulation and experimental datasets, as shown in Fig. \ref{fig:multiscale_radonpy}. The empirical generalization errors for both types of data show a monotonic decreasing trend. In particular, the increase in the size of the experimental dataset results in a significantly larger gain than the increase in simulation dataset size. The power-law curve in Eq. \eqref{eq:scale_nm} was estimated as follows:
\begin{align}
\label{eq:2Dcurve}
R(n, m) = (0.0192 + 0.338n^{-0.265}) \, m^{-0.239} + 0.0535 
\end{align}
As predicted theoretically, the scaling effect on the simulation data weakens as the experimental dataset size increases. Likewise, it was confirmed that the scaling effect on experimental data also decreases as the amount of simulation data increases.

Furthermore, by applying the concept of a marginal rate of substitution from microeconomics \citep{varian2014intermediate} to this estimated surface, we estimated the number of simulation samples equivalent to one experimental sample. For example, at the current maximum sample sizes of $n=71,\!068$ and $m= 601$, the partial derivatives of the estimated error function, analogous to marginal utility in economic theory, are given as follows:
\begin{align*}
 \frac{\partial R}{\partial n} \bigg|_{n=71068, m=601} = -1.41 \times 10^{-8}
 , \hspace{12pt}
  \frac{\partial R}{\partial m} \bigg|_{n=71068, m=601} = -3.13 \times 10^{-6}.
\end{align*}
Taking the ratio of these coefficients provides an estimate of the marginal rate of substitution between experiments and simulations. Specifically, on the set of $(m,n)$ pairs that maintain the same level of generalization error $R(n,m) = r$ (referred to as indifference curves in microeconomics), the marginal rate of substitution $\mathrm{d} m / \mathrm{d}n $ is given by $\frac{\mathrm{d} m}{\mathrm{d}n} = - \frac{\partial R}{\partial n}/\frac{\partial R}{\partial m}$ (see the Method section). In this case, 221 simulation samples are equivalent to one experimental sample.

\begin{figure}[ht]
    \centering
    \includegraphics[clip, width=0.7\columnwidth]{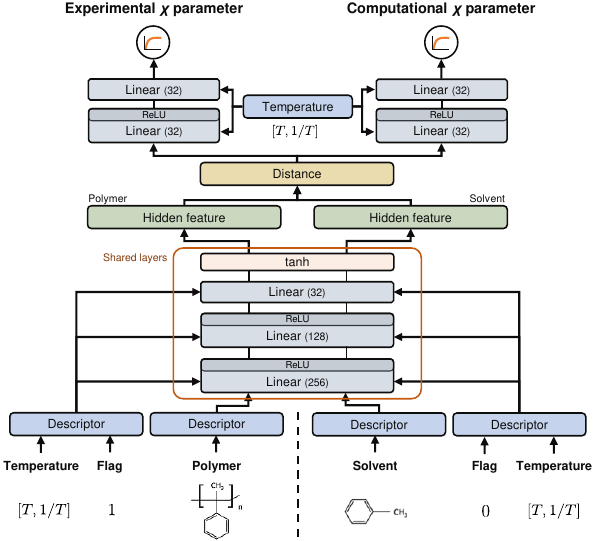}
    \caption{
        Model architecture of Sim2Real multitask learning used for predicting the Flory--Huggins $\chi$ parameter. 
    }
    \label{fig:chi1}
\end{figure}

\subsection*{Sim2Real multitask learning for polymer--solvent miscibility}
\label{sec:chi}

While the theoretical implications presented by \citet{Mikami2021ASL} were derived under the assumption of neural fine-tuning, here we explored the scaling behavior for Sim2Real multitask learning scenarios. The task is to predict the Flory--Huggins $\chi$ parameter between any given polymer and solvent, which is a critical dimensionless quantity governing the miscibility of polymer--solvent binary mixtures. The dataset comprises $\chi$ parameters for $9,\!575$ polymer--solvent pairs calculated via COSMO-RS simulations based on density functional calculations \citep{loschen2014prediction}, which were generated in our previous work \citep{aoki2023multitask}, and 1,190 experimentally observed $\chi$ parameters for 766 unique polymer--solvent pairs compiled from \citet{Orwoll-Arnold2007}. \citet{aoki2023multitask} demonstrated that integrating both simulated and experimental $\chi$ parameters into multitask learning significantly enhanced the generalization capability of the resulting predictors for experimental $\chi$ parameters. In particular, this strategy effectively addressed limitations of molecular diversity, data size constraints, and inherent distributional biases in the experimental dataset.

We slightly modified the model structure developed by \citet{aoki2023multitask} as the multitasking network architecture inspired by domain knowledge, known as the Hansen solubility parameter, as shown in Fig. \ref{fig:chi1}. The 325-dimensional descriptor encodes the chemical structure of a given polymer repeating unit or solvent. Specifically, it comprises a 190-dimensional kernel mean force field descriptor \citep{kusaba2023representation} and a 135-dimensional RDKit descriptor \citep{rdkit}, where irrelevant features with zero variance within the given dataset were removed from the 207-dimensional RDKit descriptor. Additionally, we included a binary flag representing polymer (1) or solvent (0), temperature $T$, and its inverse $1/T$ as additional inputs. For a given polymer or solvent, the input descriptor was passed through three hidden layers to map it to a 32-dimensional latent space. The distance between the polymer and solvent in this latent space was calculated, and two separate head networks were employed to output the experimental and simulated $\chi$ parameters. See the Methods section for further details on the model structure.

\begin{figure}[!t]
    \centering
    \includegraphics[clip, width=\columnwidth]{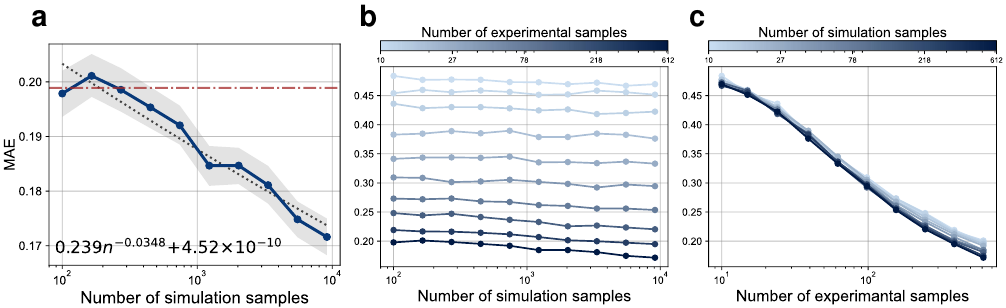}
    \caption{
        Scaling law observed in the Flory--Huggins $\chi$ parameter prediction task. 
        (a) Scaling behavior when increasing the simulation dataset size. The horizontal axis represents the number of polymer--solvent pairs used as the simulation dataset, and the vertical axis shows the average MAE of 100 independent trials with 90\% confidence interval calculated via bootstrapping. The dashed line is the estimated power-law with the estimated equation given at the bottom left, and the horizontal red line indicates the average MAE for direct learning without pretraining.
        (b) Scaling behaviors across different sizes of experimental data, and (c) scaling to increase the experimental dataset for different simulation dataset sizes. Each line shows the average MAE over 100 trials.
    }
    \label{fig:chi2}
\end{figure}

To assess the generalization performances, 20\% of the experimental data was randomly allocated as a test set, while the remaining samples, along with $n$ randomly selected simulation data points, were used for model training. This procedure was repeated 100 times independently with different data splits. The $n$ ranged across 10 evenly spaced points on a logarithmic scale within the interval $[100, 9575]$.

Fig. \ref{fig:chi2}a shows the observed scaling curves, which exhibit strong linear decay in the generalization error on a logarithmic scale. The estimated $C$ was $4.52 \times 10^{-10}$, suggesting that expanding COSMO-RS simulations can yield high-performance predictors for real systems. This observation suggests that scaling behavior occurs even in multitask learning, although the Sim2Real scaling law was theoretically derived for fine-tuning scenarios in \citet{Mikami2021ASL}.

Here, we discuss the multidimensional scaling of multitask learning. Figs. \ref{fig:chi2}b-c. describe the observed scaling behaviors when simultaneously varying the sizes of simulation and experimental datasets. Interestingly, unlike the fine-tuning results shown in the previous section, in multitask learning, as the experimental dataset size increases, the absolute gradient of the scaling curve becomes steeper. Similarly, by increasing the simulation dataset size, the improvement in generalization performance per experimental sample also increases. In other words, the experimental results suggest a mechanism where simulations and experiments mutually enhance their impact on improving generalization performance through synergistic effects. 

This observation differs from the theoretical implication of Eq. \eqref{eq:scale_nm}. This is thought to be due to the difference in the choice of fine-tuning and multitask learning. Consequently, the parameter estimation for two-dimensional scaling with Eq. \eqref{eq:scale_nm} is invalid. Instead, the equivalent sample size was calculated by approximating the gradient based on the observed increment of the MAE for increasing simulation and experimental dataset sizes. The estimated scaling curves and the gradients at the current dataset sizes ($n=9,\!129$ and $m=612$) were computed respectively as follows:
\begin{align*}
    R(n, 612) &= 0.239 n^{-0.0348} + 4.52\times 10^{-10}, \hspace{12pt} \frac{\partial R(n ,612)}{\partial n} \bigg|_{n=9129} = -6.63 \times 10^{-7}, \\
    R(9129, m) &= 0.863 n^{-0.240} + 6.36\times 10^{-16}, \hspace{12pt} \frac{\partial R(9129 ,m)}{\partial m} \bigg|_{m=612} = -7.25 \times 10^{-5}.
\end{align*}
By taking the ratio of these gradients, the marginal rate of substitution was estimated to be 109, indicating that for the $\chi$ parameter prediction task, one real-world experiment is worth 109 COSMO-RS simulations.

\begin{figure}[!t]
    \centering
    \includegraphics[clip, width=\columnwidth]{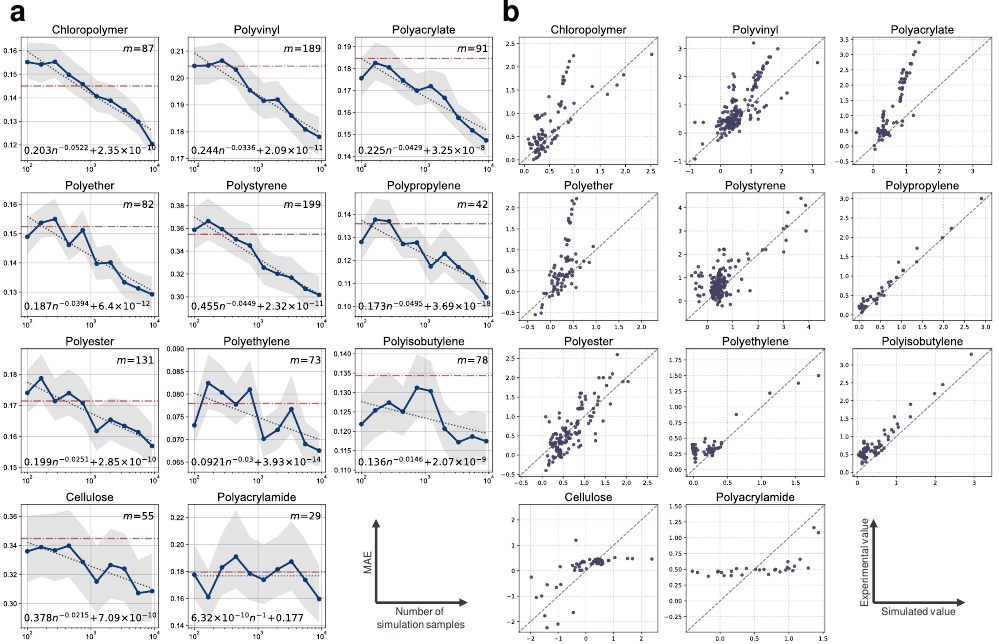}
    \caption{
    (a) Observation of Sim2Real scaling behaviors for different polymer classes in the $\chi$ parameter prediction task. Test instances of polymer--solvent pairs were classified into 11 classes based on structural features. The $m$ value is denoted in the upper-right corner of each panel. 
    (b) Predictive capability of COSMO-RS simulations (horizontal axis) against experimental values (vertical axis) for each of the 11 polymer classes in the $\chi$ parameter predictions. 
    }
    \label{fig:chi5}
\end{figure}

Although we investigated whether the overall generalization performance scales across various materials as a whole, it is important to verify whether individual materials scale or not. Fig. \ref{fig:chi5}a summarizes the observed scaling behaviors to increasing $n$ for different polymer classes, where test instances of polymer--solvent pairs were classified into 11 classes based on structural features of the polymers. The generalization performances of chloropolymers, polyvinyls, polyacrylates, polyethers, and polystyrenes were strongly scaled, while, other materials showed almost no improvement, e.g., polyacrylamides. Observing the scalability of each material class provides valuable insights for planning data generation. In the development of a simulation database, limited computational resources should be allocated primarily to scalable polymer classes. For non-scalable polymer classes, some modifications are needed in the data production protocol to improve scalability. 
To devise strategies, it is important to identify the governing factor that determines the scalability. Fig. \ref{fig:chi5}b shows parity plots of $\chi$ parameters obtained from COSMO-RS simulations and experimental values for each polymer class. In comparison with the scaling behaviors shown in Fig. \ref{fig:chi5}a, it is evident that the observed scalability of polymer species can be largely explained by the predictive capability of COSMO-RS simulations. However, in some polymer classes such as polystyrenes and polyesters, where the predictive ability of COSMO-RS simulations is weak, the generalization performance of transferred models scales strongly, reaching levels far beyond those of the simulations. This indicates the essence of Sim2Real transfer. Additionally, it is important to note that many of the non-scalable polymer classes had extremely limited experimental data (see Fig. \ref{fig:chi5}a). For example, the $m$ values for celluloses and polyacrylamides are 55 and 29, respectively. In such cases, model training becomes extremely challenging. Furthermore, empirical generalization errors approximated with the small sample sets may significantly deviate from true generalization errors. Even if a polymer class appears to be non-scalable, it cannot be conclusively determined that there is no transferability or scalability. 

\subsection*{Transfer learning for thermal and electrical conductivity of inorganic materials}
\label{sec:wf}

\begin{figure}[ht]
    \centering
    \includegraphics[clip, width=\columnwidth]{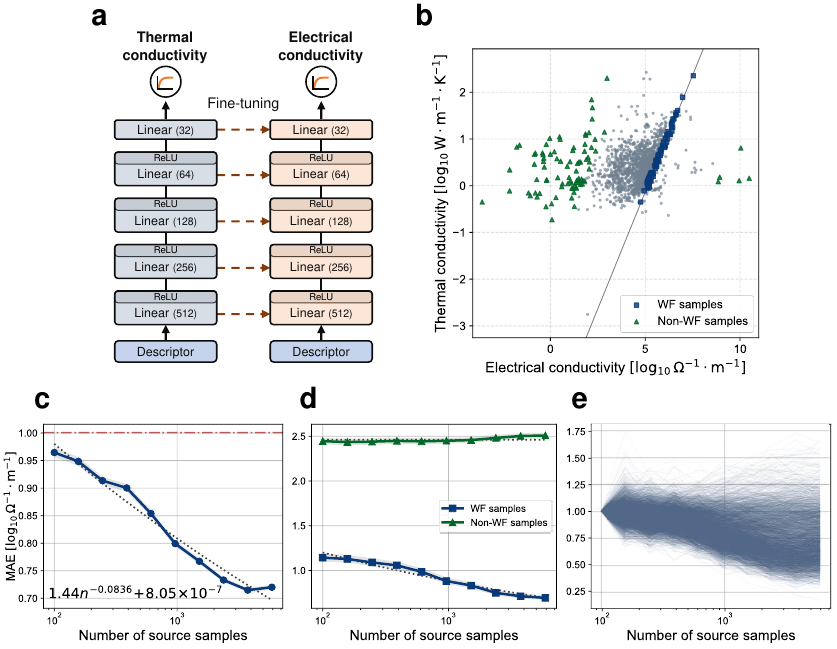}
    \caption{
        Real2Real transfer learning from thermal to electrical conductivities. (a) Model architecture. (b) Parity plot showing values of electrical conductivity (horizontal axis) and thermal conductivity (vertical axis) at 300 K on a logarithmic scale. The dashed line represents the WF law. Blue square dots represent the top 10\% of compounds with the smallest deviation from the WF rule (WF samples), while green triangular dots correspond to the top 10\% of compounds with the largest deviation (non-WF samples). (c) Scaling behavior. The horizontal axis represents the source data size of the electrical conductivity prediction task, while the vertical axis shows the average MAE for the thermal conductivity prediction task over 100 independent trials (solid line) with the standard deviation and 90\% confidence interval calculated using bootstrapping. The black dashed line represents the estimated power-law (equation provided at the bottom left), and the red dashed line indicates the MAE for direct learning. (d) Scaling behavior for the two extracted datasets. The line color corresponds to the color of the dots in the parity plot in (b). (e) Scaling behavior for each of the 3,640 compounds in the dataset.
    }
    \label{fig:WF1}
\end{figure}

By definition, the scaling laws of transfer learning hold not only for Sim2Real scenarios but also for real-to-real (Real2Real) transfer scenarios. Here, we highlight an interesting aspect of the scaling analysis by showing the Real2Real scaling behavior in transfer learning from thermal conductivity to electrical conductivity for inorganic compounds.

We compiled a dataset from Starrydata \citep{katsura2019data}, comprising $5,\!910$ inorganic compounds with experimentally observed thermal conductivities and $3,\!640$ compounds with experimental electrical conductivities, all derived at 300 K. Starrydata is a comprehensive experimental database of thermoelectric materials that was collected from published papers. Fig. \ref{fig:WF1}b illustrates the dependency and discrepancy between the two physical properties across $1,\!757$ materials, where both thermal and electrical conductivity measurements were obtainable. According to the WF law \citep{jones1985theoretical}, the ratio of thermal conductivity ($\kappa$) to electrical conductivity ($\sigma$) of a metal is proportional to temperature ($T$), expressed as:
\[
    \frac{\kappa}{\sigma} = LT,
\]
where $L=2.44 \times 10^{-8} \mathrm{W \Omega} \mathrm{K}^{2}$ is the Lorentz number. The gray line in Fig. \ref{fig:WF1}b depicts the WF law on the joint distribution of thermal and electrical conductivities at 300 K. While the WF law holds for metallic materials, where the free electrons are mainly responsible for both of these properties, it does not necessarily hold for non-metallic materials. Since the data includes both metallic and non-metallic materials, some of the samples deviated from this line. 

For model building, we encoded the compositional features of an input compound into a 580-dimensional kernel mean descriptor \citep{kusaba2023representation}. Subsequently, a fully connected neural network was pretrained to learn the mapping from the vectorized composition to thermal conductivity. The network architecture is illustrated in Fig. \ref{fig:WF1}a. The $n$ used to train the thermal conductivity predictor was increased logarithmically in 10 steps from 100 to $5,\!910$. During the transfer learning phase, 80\% of the electrical conductivity data were randomly selected for fine-tuning, and the remaining 20\% served as the test set to evaluate the performance of the transferred model. This procedure was repeated 500 times with different randomly selected sample sets.

Fig. \ref{fig:WF1}c shows the observed scaling behaviors. The predictive performance improved linearly on the logarithmic scale, with the estimated power-law function $1.44n^{-0.0836} + 8.05 \times 10^{-7}$. Since the WF law holds for metallic materials, the transfer is expected to be more successful for metallic materials than for non-metallic ones. To investigate the difference in the transferability for metallic and non-metallic materials, we extracted samples that followed the WF law (blue square dots in Fig. \ref{fig:WF1}b) and those that deviate from it (green triangular dots in Fig. \ref{fig:WF1}b ). Fig. \ref{fig:WF1}c shows the scaling behaviors separately for each of the two sample sets. As expected, strong scaling was observed for materials for which the WF law holds, but for non-metallic materials that deviate from this law, increasing the amount of thermal conductivity data did not improve predictive performance. 

Furthermore, observing transferability individually for different materials allows us to infer the presence or absence of common physical mechanisms between different physical systems. Fig. \ref{fig:WF1}e illustrates the scaling behaviors of all test cases, which clearly distinguishes material species where transfer does or does not scale. The presence or absence of scaling laws and the observed scaling strength could be used to characterize individual materials. Moreover, observing individual transferability provides valuable insights for planning data generation. The overall average performance transitioned into a plateau around $n=4 \times 10^{3}$ (Fig. \ref{fig:WF1}c). However, there were several material groups where predictive performance continued to improve logarithmically; for example, metallic materials. (Figs. \ref{fig:WF1}d-e). Intuitively, it would be efficient to halt the production of source data for material groups where improvement has plateaued and reallocate resources to groups more likely to scale. Analyzing only the overall average generalization performance overlooks the existence of material groups with the potential to scale even further.

\section*{Discussion}\label{sec12}
This study discussed the significance and utility of analyzing the scalability in Sim2Real and Real2Real transfer learning in materials science. Across diverse case studies encompassing polymers and inorganic materials, it was consistently observed that as the size of the computational pretraining data set increases, the prediction error relative to the experimental data improves according to a power-law relationship. These findings highlight the importance of synergistic effects between computational and experimental approaches. By observing the scaling law for Sim2Real transfer, we can estimate the required size of computational datasets to achieve the desired predictive performance in downstream real-world tasks. Additionally, we provide a microeconomic framework for determining the optimal allocation of computational and experimental resources during the creation of data platforms by analyzing multi-dimensional scaling behaviors. This approach guides decisions related to the allocation of resources for data collection efforts for maximum impact.

The scaling laws of transfer learning provide guiding principles for designing computational databases. It is desirable to create transferable computational databases that scale the generalization performance of downstream tasks for specified target tasks in the real-world domain. Alternatively, it is important to discover real-world tasks and analytical workflows that can be transferred scalably from computational databases. While various computational material-property databases have been developed to date, there are no reported cases of the values being quantified from the perspective of scaling laws. Strong scalability of transfer to diverse real-world tasks serves as a measure of the usefulness of the computational database. 

It is important to see that discrepancies always exist between simulated and experimental properties. Additionally, experimental data are subject to biases and fluctuations due to unobserved factors related to the experimental conditions, sample fabrication, noise in measurement systems, and selection bias of the researchers. Therefore, transfer learning plays a key role in bridging the gap between complex and uncertain real-world scenarios and imperfect computational models. To this end, it is crucial to explicitly demonstrate the transferability and benefits of expanding datasets to downstream tasks. Finding a scheme with the scalability of Sim2Real transfer is a goal of developing materials databases using simulated data.

\section*{Methods}

\subsection*{PoLyInfo polymer property datasets}
Experimental property datasets for refractive index, density, $C_{\mathrm{P}}$, and thermal conductivity were extracted from the polymer property database PoLyInfo \citep{polyinfo_paper, polyinfo_paper2, polyinfo_web}. The data for density, specific heat capacity, and thermal conductivity were restricted to measurements in amorphous states near room temperature (273-323K). The number of polymers in each property dataset was 39 for thermal conductivity, 104 for $C_{\mathrm{P}}$, 607 for density, and 234 for refractive index.

\subsection*{RadonPy polymer property datasets}

To construct the simulation datasets, all-atom classical MD simulations were conducted using RadonPy, a Python library that automates polymer property calculations through high-throughput MD simulations \citep{hayashi2022radonpy}. Input parameters include the chemical structure of polymer repeating units represented by a simplified molecular input line entry system (SMILES) \citep{weininger1988smiles}, polymerization degree, number of polymer chains forming a simulation cell, temperature, and pressure. The automated calculation workflow consists of the following steps: (1) conformation search for a monomer with the given repeating unit, (2) calculation of electronic properties such as atomic charges using the density functional theory (DFT) method, (3) search for initial configuration of polymer chains using the self-avoiding random walk algorithm, (4) assignment of force field parameters using the general Amber force field version 2 (GAFF2), (5) generation of isotropic amorphous cells, (6) MD simulations for equilibration, (7) determination of whether to reach equilibrium, (8) non-equilibrium MD (NEMD) simulations for thermal conductivity calculation, and (9) property calculation in the post-processing stage. The DFT calculations and the MD simulations were executed using Psi4 \citep{smith2020psi4} and LAMMPS, respectively, within the RadonPy interface. An amorphous cell was created with 10 polymer chains comprising approximately $10,\!000$ atoms. Following the initial configuration of polymer chains using the self-avoiding random walk and a 1 ns NVT simulation, the simulation cell was packed isotropically to achieve a density of 0.8 $\mathrm{g} \cdot\mathrm{cm}^{-3}$ at 700 K. The amorphous cell was equilibrated following Larsen's 21-step compression/decompression equilibration protocol \citep{larsen2011molecular}, undergoing temperature cycles between 300 and 600 K, repeating the ascent and descent for stabilization. After completing the 21-step equilibration process, NpT simulations were conducted for over 5 ns at 300 K and 1 atm until reaching equilibrium. The property calculation methods for the density, specific heat capacity at constant pressure, refractive index, and thermal conductivity were described previously \citep{hayashi2022radonpy} and detailed in the Supplementary Information.

In collaboration with an academia--industry consortium, we generated the property datasets of approximately $7 \times 10^4$ linear polymers in amorphous states using RadonPy (see Table S1). The virtual polymers were generated using an N-gram-based polymer structure generator \citep{ikebata2017bayesian} for each of the 20 polymer classes, such as polyimides, polyesters, and polystyrenes, following the classification rule established by PolyInfo. The chemical structure $X$ of an existing compound used in the training dataset is described by the SMILES representation, where $X$ is represented by a string of length $p$ as $X=x_1 x_2 \ldots x_p$. By using the string set of synthesized polymers belonging to each polymer class, an N-gram language model was trained to obtain a structure generator that mimicked the patterns, such as frequent fragments and appropriate chemical bonding rules, observed for the existing polymers. The 20 class-specific SMILES generators were used to create the virtual library. A list of the 20 polymer classes with their dataset sizes is provided in Table S1.

\subsection*{Equivalent sample size for experimental and simulation data}

Differentiating the generalization error $R(n,m)$ in Eq. \eqref{eq:scale_nm} with respect to $n$ and $m$, we obtain the following expression:
\begin{eqnarray}
\mathrm{d} R(n,m) = \frac{\partial R}{\partial n} \mathrm{d} n + \frac{\partial R}{\partial m} \mathrm{d} m.
\end{eqnarray}
On the set of equivalent samples $(n, m)$ that maintain the same level of $R(n, m) = r$, $R(n,m)$ remains constant at $r$, thus satisfying $\mathrm{d} R(n,m) =0$. Therefore, $\frac{\mathrm{d} m}{ \mathrm{d}n} = - \frac{\partial R}{\partial n}/\frac{\partial R}{\partial m}$ holds.

\subsection*{Polymer--solvent solubility datasets}
\citet{aoki2023multitask} used the experimental values of the $\chi$ parameter for $1,\!190$ polymer--solvent pairs, consisting of 46 different polymers and 140 different solvent molecules, to train the model. 
 The data were compiled from a supplementary table of \citet{Orwoll-Arnold2007}. The dataset also included measurements of the $\chi$ parameter for different temperatures and polymer--solvent compositions. The molecular species of the polymers/solvents in the dataset were distributed over a limited region of the entire chemical space. In addition, in certain experimental systems, it is difficult to measure the $\chi$ parameters of the polymer--solvent system in an immiscible state, resulting in a significant bias in the distribution of the data. Therefore, models trained using only this dataset generally have narrow predictive applicability.

To tackle this issue, we utilized the COSMO-RS simulation \cite{KlamtCOSMO1,KlamtCOSMO2,KlamtCOSMO3,Klamt2005} to generate a dataset of $\chi$ parameters for $9,\!129$ pairs of polymers and solvents at the BP-SVP-AM1 level \cite{aoki2023multitask}. The calculations were performed using the TURBOMOLE \cite{TURBOMOLE} and COSMOtherm \cite{COSMOtherm} software packages for creating COSMO files by density functional calculations and the calculations of $\chi$ parameters from the COSMO files, respectively. For polymers, a structure comprising three repeating units was created, in which the two endpoints were replaced by methyl groups. After creating the COSMO files, the COSMOmeso function was executed to calculate the $\chi$ parameters using the activity coefficients obtained from the COSMO files. 

\subsection*{Data preprocessing}

In all experiments, variable transformations were applied to the model inputs and outputs to enhance the efficiency of machine-learning model training. The methods included logarithmic transformation, normalization, Yeo--Johnson transformation \citep{yeo2000new}, and min--max transformation (i.e., scaling each feature to a range of $[0, 1]$). These methods were tailored for the input and output variables in each of the three applications, as summarized in Table \ref{tbl:data}.

\begin{table}[t]
    \caption{
        Data preprocessing methods for input and output variables in three different applications: polymer property prediction using RadonPy (RadonPy), multitask learning of polymer--solvent miscibility ($\chi$ parameter), and transfer learning between thermal and electrical conductivities using Starrydata (Starraydata). A combination of four methods --- logarithmic transformation (log), normalization (norm), Yeo--Jhonson transformation (YJ), and min--max transformation (0-1) --- was applied in the order described below.
    }
    \begin{tabular*}{\textwidth}{@{\extracolsep\fill}llll}
    \toprule%
    & RadonPy & $\chi$ parameter & Starrydata \\
    \midrule
    \multirow{2}{*}{Input} & \multirow{2}{*}{log $\rightarrow$ norm $\rightarrow$ YJ} & FFKM\footnotemark[1]: log $\rightarrow$ norm $\rightarrow$ YJ $\rightarrow$ 0-1 & \multirow{2}{*}{log $\rightarrow$ norm $\rightarrow$ YJ} \\
     & & RDKit + temp + inv-temp\footnotemark[2]: 0-1 & \\
    Output  & norm $\rightarrow$ YJ & norm $\rightarrow$ YJ & norm $\rightarrow$ YJ $\rightarrow$ 0-1 \\
    \botrule
    \end{tabular*}
    \footnotetext[1]{Force-field kernel mean (FFKM) descriptor \citep{kusaba2023representation}.}
    \footnotetext[2]{Concatenation of RDKit descriptor \citep{rdkit} (RDKit), thermodynamic temperature (temp), and inverse temperature (inv-temp).}
    \label{tbl:data}
\end{table}

\subsection*{Model fine-tuning}
In Sim2Real transfer learning using RadonPy and transfer learning between thermal and electrical conductivities, neural fine-tuning was employed. Specifically, the weights of the neural networks pretrained on the source data (MD-calculated property data or experimental data for thermal conductivity) were used as initial values and updated with the target data (experimental observation of polymeric properties or electrical conductivity data). 
The hyperparameters, such as learning rate and batch size, are listed in Table \ref{tbl:hpara}.

\subsection*{Multitask learning}
For the $\chi$ parameter prediction task, we employed multitask learning with empirical risk minimization as follows:
\begin{align*}
    \min_f \frac{\lambda}{|\mathcal{D}_{\text{sim}}|} \sum_{(\chi, T, p, s) \in \mathcal{D}_{\text{sim}}} \bigl\{ \chi - f(T, p, s) \bigr\}^2
    + \frac{1-\lambda}{|\mathcal{D}_{\text{exp}}|} \sum_{(\chi, T, p, s) \in \mathcal{D}_{\text{exp}}} \bigl\{ \chi - f(T, p, s) \bigr\}^2,
\end{align*}
where $\mathcal{D}_{\text{sim}}$ and $\mathcal{D}_{\text{exp}}$ denote the dataset of $\chi$ parameters obtained by the COSMO-RS simulation and the experimental dataset, respectively. 
The neural network model $f(T, p, s)$ is a function of temperature $T$, polymer $p$, and solvent $s$.
The first term fits the simulated dataset, while the second term fits the experimentally observed dataset. The hyperparameter $\lambda$ controls the relative importance between these two terms. In this study, we set $\lambda=0.5$, which is consistent with the value employed in \citet{aoki2023multitask}, resulting in learning from both simulation and real systems with equal importance.
Other hyperparameters are listed in Table \ref{tbl:hpara}.

\begin{table}[t]
    \caption{
        Hyperparameter settings for model training in the three applications: RadonPy, $\chi$ parameter, and Starrydata. 
        In the fine-tuning experiments (RadonPy and $\chi$ parameter), the same hyperparameters were used for the source and target tasks, except for the learning rate in Starrydata.
    }
    \begin{tabular*}{\textwidth}{@{\extracolsep\fill}llll}
    \toprule%
    & RadonPy & $\chi$ parameter & Starrydata \\
    \midrule
    Optimizer & Adam \citep{kingma2014adam} & Adam \citep{kingma2014adam} & Adam \citep{kingma2014adam} \\
    \multirow{2}{*}{Learning rate} & \multirow{2}{*}{$0.001$} & \multirow{2}{*}{$0.001$} & $0.001$ (source task) \\
     & & & $0.0001$ (target task) \\
    Batch size & $32$ & $16$ & $32$ \\
    Early stopping patience & $5$ & $10$ & $10$ \\
    \botrule
    \end{tabular*}
    \label{tbl:hpara}
\end{table}

\backmatter

\section*{Data availability}
The data supporting the findings of this study will be made available upon reasonable request to the corresponding author. The datasets of the experimental and computational $\chi$ parameters can be accessed via Figshare \url{https://github.com/yoshida-lab/MTL_ChiParameter}. The datasets of thermal and electrical conductivity are accessible through Starrydata \url{https://figshare.com/projects/Starrydata_datasets/155129}. 

\section*{Code availability}
The code for multitask learning for the $\chi$ parameter prediction task is available at the Github \url{https://github.com/yoshida-lab/MTL_ChiParameter}. Other codes are available upon reasonable request to the corresponding author. 

\section*{Acknowledgements}
We express our sincere gratitude to all members of ISM–MCC Frontier Materials Design Laboratory, a joint laboratory of Mitsubishi Chemical Corporation (MCC) and the Institute of Statistical Mathematics (ISM), for their valuable contributions to the discussion of this study. This research received support from MEXT as ``Program for Promoting Researches on the Supercomputer Fugaku'' (project ID: hp210264), JST CREST (Grant Numbers JPMJCR19I3, JPMJCR22O3, JPMJCR2332), MEXT/JSPS KAKENHI Grant-in-Aid for Scientific Research on Innovative Areas (19H05820), Grant-in-Aid for Scientific Research (A) (19H01132), Grant-in-Aid for Research Activity Start-up (23K19980), and Grant-in-Aid for Scientific Research (C) (22K11949). Computational resources were provided by Fugaku at the RIKEN Center for Computational Science, Kobe, Japan (hp210264) and the supercomputer at the Research Center for Computational Science, Okazaki, Japan (project: 23-IMS-C113, 24-IMS-C107).

\section*{Author contributions}

R.Y. and S.M. devised the project, main conceptual ideas, and outline proof. S.M. and Y.H. implemented the machine-learning algorithms and conducted the experiments with the support of R.Y., S.W., H.S., and K.S. Y.H. performed the MD simulations using RadonPy to generate the polymer property data. K.S. generated the $\chi$ parameter dataset using the COSMO-RS simulations. K.F. performed a theoretical analysis of the Sim2Real transfer learning. H.S. examined the results from a physicochemical point of view. M.I. and I.K. extracted and structured data from PoLyInfo. S.M. and R.Y. wrote the manuscript.

\section*{Competing interests}
The authors declare no competing interests.

\bibliography{sn-bibliography}

\end{document}


\title[Scaling law of Sim2Real transfer learning in materials science]{
\centering {\large Supplementary Information} \\[1ex] 
Scaling Law of Sim2Real Transfer Learning in Expanding Computational Materials Databases for Real-World Predictions}

\author*[1]{\fnm{Shunya} \sur{Minami}}\equalcont{These authors contributed equally to this work.}
\author[1,2]{\fnm{Yoshihiro} \sur{Hayashi}}
\equalcont{These authors contributed equally to this work.}
\author[1,2]{\fnm{Stephen} \sur{Wu}}
\author[1,2]{\fnm{Kenji} \sur{Fukumizu}}
\author[3]{\fnm{Hiroki} \sur{Sugisawa}}
\author[4]{\fnm{Masashi} \sur{Ishii}}
\author[4]{\fnm{Isao} \sur{Kuwajima}}
\author[3]{\fnm{Kazuya} \sur{Shiratori}}
\author*[1,2,4]{\fnm{Ryo} \sur{Yoshida}}\email{yoshidar@ism.ac.jp}
\affil[1]{\orgdiv{The Institute of Statistical Mathematics}, \orgname{Research Organization of Information and Systems}, \orgaddress{\city{Tachikawa}, \postcode{190-8562}, \country{Japan}}}
\affil[2]{\orgdiv{The Graduate Institute for Advanced Studies}, \orgname{SOKENDAI}, \orgaddress{\city{Tachikawa}, \postcode{190-8562}, \country{Japan}}}
\affil[3]{\orgdiv{Science \& Innovation Center}, \orgname{Mitsubishi Chemical Corporation}, \orgaddress{\city{Yokohama}, \postcode{227-8502}, \country{Japan}}}
\affil[4]{\orgdiv{Research and Service Division of Materials Data and Integrated System}, \orgname{National Institute for Materials Science}, \orgaddress{\city{Tsukuba}, \postcode{305-0047}, \country{Japan}}}

\maketitle

\section*{Polymer property calculation using RadonPy}
RadonPy, available as an open-source Python library (\url{https://github.com/RadonPy/RadonPy}), seamlessly automates equilibrium MD and non-equilibrium MD (NEMD) simulations for diverse polymers and their associated physical properties. The simulation engine integrates LAMMPS with the ab initio quantum chemistry program Psi4, while GAFF2 parameterizes organic polymers. Input/output systems are implemented using the Python chemoinformatics library RDKit.

For linear polymers in amorphous states, RadonPy requires input parameters such as a SMILES string, polymerization degree, number of polymer chains, and temperature. In this study, an amorphous cell was constructed, comprising 10 polymer chains with around 10,000 atoms. The workflow included an automated conformation search, DFT-based electronic property calculations, initial configuration generation, force field parameter assignment, isotropic cell creation, equilibration MD simulation, equilibrium status assessment, NEMD execution for thermal conductivity, and physical property calculations.

In this study, we used a property dataset comprising approximately 70,000 amorphous polymers, including density, specific heat capacity at constant pressure, refractive index, and thermal conductivity, produced on the Fugaku supercomputer. This dataset was collaboratively developed with the RadonPy consortium, including approximately 240 participants from national institutes, 9 universities, and 36 companies. 

The density was computed using the mass $m$ and volume $V$ of the system in equilibrium MD with the NpT ensemble as follows:
\begin{eqnarray}
\rho = \frac{m}{\langle V \rangle}
\end{eqnarray}
where the angular brackets denote time averaging.

The specific heat capacity at constant pressure $C_{\mathrm{P}}$ was calculated from the fluctuations of the enthalpy $H$ in equilibrium MD with the NpT ensemble:
\begin{eqnarray}
C_{\mathrm{P}} = \frac{\langle \delta H \rangle}{k_{\mathrm{B}} T^2 m}
\end{eqnarray}
where $ k_{\mathrm{B}}$ is the Boltzmann constant, and $T$ is the temperature. The enthalpy was calculated under a constant pressure of 1 atm.

The refractive index $n$ was derived from the Lorentz--Lorenz equation:
\begin{eqnarray}
\frac{n^2 - 1}{n^2 + 2}= \frac{ 4 \pi}{3}\frac{\rho}{M}\alpha_{\mathrm{polar}}
\end{eqnarray}
Here, $\alpha_{\mathrm{polar}}$ is the isotropic dipole polarizability of a repeating unit calculated from the DFT calculation, and $M$ is the molecular weight of a repeating unit.

To calculate the thermal conductivity, we performed the reverse NEMD simulation proposed by \citet{muller1997simple}. The thermal conductivity $\kappa$ was calculated according to Fourier’s law:
\begin{eqnarray}
\kappa = \frac{J_Q}{\partial T/ \partial x} = \frac{\Delta E}{2A \Delta t (\partial T/ \partial x)}
\end{eqnarray}
where $ J_Q $ is the heat flux, and $\partial T/ \partial x $ is the temperature gradient of the NEMD simulation. The heat flux was calculated from the exchanged energy $\Delta E$, the cross-sectional area in the heat flux direction $A$, and the simulation time $\Delta t $.

\section*{RadonPy datasets}
The density, specific heat capacity at constant pressure, refractive index, and thermal conductivity of approximately $7 \times 10^4$ amorphous polymers were calculated using RadonPy, as described above and in the Method section. Details of the MD simulations are fully described in \citet{hayashi2022radonpy}. The virtual polymers were generated using an N-gram-based polymer structure generator \citep{ikebata2017bayesian} for each of the 20 polymer classes, including polyimides, polyesters, and polystyrenes, following the classification rule established by PolyInfo. Using the SMILES string set of synthesized polymers belonging to each polymer class in PoLyInfo, an N-gram language model was trained to obtain a structure generator that mimicked the patterns (frequent fragments, appropriate chemical bonding rules, etc.) for the existing polymers. The 20 class-specific SMILES generators were used to create the virtual library. The list of the 20 polymer classes with their data sizes is provided in Table \ref{tab:polymer_class}.

\begin{table}
    \centering
    \caption{Summary of RadonPy datasets}
    \begin{tabular}{ccccc}
        \toprule
        Polymer class \tnote{1} & Density & $C_{\mathrm{P}}$ & Refractive index & Thermal conductivity \\
        \midrule
        Polyolefins (hydrocarbons) & 1723 & 1723 & 1717 & 1712 \\
        Polystyrenes & 5981 & 5918 & 5866 & 5524 \\
        Polyvinyls & 5076 & 5076 & 5042 & 4995 \\
        Polyacrylates & 4613 & 4155 & 4120 & 4019 \\
        Polyhalo-olefins & 1331 & 1331 & 1324 & 841 \\
        Polydienes & 693 & 693 & 687 & 636 \\
        Polyethers & 4568 & 4373 & 4362 & 4273 \\
        Polysulfides & 3693 & 3692 & 3680 & 3397 \\
        Polyesters & 4190 & 4190 & 4173 & 4157 \\
        Polyamides & 3563 & 3563 & 3476 & 3543 \\
        Polyurethanes & 2939 & 2939 & 2925 & 2457 \\
        Polyureas & 2857 & 2856 & 2582 & 2687 \\
        Polyimides & 5571 & 5570 & 5547 & 5281 \\
        Polyanhydrides & 3374 & 3374 & 3326 & 3126 \\
        Polycarbonates & 3235 & 2807 & 2803 & 2481 \\
        Polyimines & 3875 & 3871 & 3862 & 3552 \\
        Polyphosphazenes & 2915 & 2915 & 2842 & 2368  \\
        Polyketones & 4166 & 4166 & 4123 & 3843 \\
        Polysulfones & 3974 & 3973 & 3927 & 3863 \\
        Polyphenylenes & 2794 & 2794 & 2790 & 2076 \\
        \hline
        Total & 71068 & 69980 & 69174 & 64831\\
        \bottomrule
    \end{tabular}
    \begin{tablenotes}
        \item[1] A polymer can belong to multiple classes.
    \end{tablenotes}
    \label{tab:polymer_class}
\end{table}

\bibliography{sn-bibliography}